\documentclass[journal]{IEEEtran}

\usepackage{amsmath}
\usepackage{float}
\usepackage{graphicx}
\usepackage{mathtools}
\usepackage[export]{adjustbox}
\usepackage[justification=centering]{caption}
\usepackage[colorlinks=true, allcolors=blue]{hyperref}
\usepackage{tikz}
\usepackage{array}
\usepackage{tabularray}
\usepackage{wrapfig}
\usepackage{enumerate}
\usepackage{tabularx}
\usepackage{subcaption}
\usepackage{multirow}
\usepackage{cite}
\usepackage{pgfplots}
\usetikzlibrary{shapes.geometric, arrows}
\usepackage{lipsum,booktabs}

\tikzstyle{startstop} = [rectangle, rounded corners, minimum width=3cm, minimum height=0.7cm,text centered, draw=black, font=\small]
\tikzstyle{io} = [trapezium, trapezium left angle=70, trapezium right angle=110, minimum width=3cm, minimum height=0.7cm, text centered, draw=black,font=\small]
\tikzstyle{process} = [rectangle, minimum width=1.5cm, minimum height=0.7cm, text centered, draw=black,font=\small]
\tikzstyle{decision} = [diamond, minimum width=3cm, minimum height=0.8cm, text badly centered, aspect=3, text badly centered, inner sep=0pt, draw=black,font=\small]
\tikzstyle{arrow} = [thick,->,>=stealth]

\usepackage{array}
\begin{document}
\title{Equal Incremental Cost-Based Optimization\\ Method to Enhance Efficiency for\\ IPOP-Type Converters
}
\author{Hanfeng Cai,
        Haiyang Liu,
        Heyang Sun, and Qiao Wang
}

\maketitle

\begin{abstract}
Systematic optimization over a wide power range is often achieved through the combination of modules of different power levels. This paper addresses the issue of enhancing the efficiency of a multiple module system connected in parallel during operation and proposes an algorithm based on equal incremental cost for dynamic load allocation. Initially, a polynomial fitting technique is employed to fit efficiency test points for individual modules. Subsequently, the equal incremental cost-based optimization is utilized to formulate an efficiency optimization and allocation scheme for the multi-module system. A simulated annealing algorithm is applied to determine the optimal power output strategy for each module at given total power flow requirement. Finally, a dual active bridge (DAB) experimental prototype with two input-parallel-output-parallel (IPOP) configurations is constructed to validate the effectiveness of the proposed strategy. Experimental results demonstrate that under the 800W operating condition, the approach in this paper achieves an efficiency improvement of over 0.74\% by comparison with equal power sharing between both modules.
\end{abstract}

\begin{IEEEkeywords}
Equal Incremental Cost, Efficiency Optimization, Load Distribution, IPOP-Connected DC-DC Converters, Dual Active Bridge (DAB).
\end{IEEEkeywords}

\IEEEpeerreviewmaketitle

\section{Introduction}
\IEEEPARstart{I}{n} this day and age, electric vehicles (EVs) \cite{EV} have become an rapidly-expanding competitor to traditional fuel cars due to their sustainability, cost-effectiveness, and reduced noise. In 2030, the market share of EVs is predicted to grow by 30\% \cite{market}, which leads to the demand of more reliable, higher power density, and low cost power electronic solutions \cite{battery_review}. Currently, there exists proposed charging infrastructure like single stage vehicle on-board chargers (OBCs) \cite{sic} and two stage AC-DC and DC-DC converter \cite{twostage}. As the global automotive industry shifts towards electric vehicles, it becomes crucial to develop ways to improve the efficiency of power converters to accommodate the diverse power and voltage requirements of different vehicles.

Systematic solutions like adaptive charging \cite{adaptive_charging} and iteration-based algorithm \cite{iteration_algorithm} have been proposed for EV charging applications. Different modulation or control strategies have been experimented \cite{soft_switching} \cite{lightload} \cite{inductive_power} to enhance efficiency, and numerous topologies like soft-switching buck-boost \cite{buck_boost} and multi-phase multi-voltage converters \cite{full_cell} are presented.  Still, one of the most effective solutions is DC-DC converters, such as Dual-Active Bridges (DAB) \cite{DAB1,DAB2,DAB3,haiyanggeiejpes}, which support bidirectional power transfer, galvanic isolation, high power density, and current-stress minimization techniques like triple-phase shift (TPS) control \cite{stress_min}.

However, increased power requirement and physical limitations requires innovations on converters. One of them is three-phase converters \cite{threephase}, \cite{sixleg}. Another proposed design utilizes input-serial-output-parallel (ISOP) configuration to improve efficiency for smart transformer applications \cite{ISOP}. To further enhance design redundancy \cite{droop_redundancy} and accommodate a wider range of power requirement \cite{two_DAB}, Input-parallel-output-parallel (IPOP)-connected converters offer a simple solution using independent modular control \cite{shedding}. Multiple converter systems are able to achieve high power rating with lower powered converters through power sharing \cite{MMDAB}. In contrast, a single module is often optimized for a power level and is unable to adjust across dynamic power requirements due to physical limitations. While some existing literature utilizes phase-shift control \cite{98.3}, soft-switching \cite{stacked_bridge}, and current sharing techniques \cite{current_sharing}. For optimization on light load conditions, dynamic active phase variation with load demand has been proposed \cite{light-load}. Another paper presents a passive current sharing method based first-order digital sine filter \cite{sine}. This article uniquely provides a algorithmic way to calculate the optimal operating conditions for each module based on their power profiles. 

\begin{figure}[H]
\centering
\includegraphics[width=0.45\textwidth]{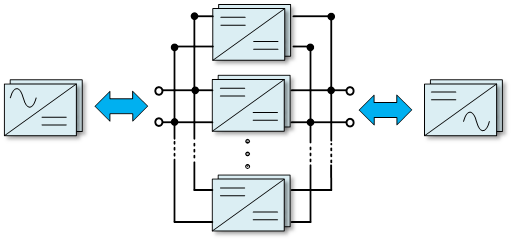}
\caption{\label{fig:multiple_converter}Schematic diagram for IPOP-connected converter modules.}
\end{figure}


This article is organized as follows: Section \ref{section:description} identifies the key problems of optimizing multiple IPOP-connected converters, Section \ref{section:solution} showcases the solutions in four steps: A curve-fitting approach for efficiency analysis, followed by priority list (PL) method and equal incremental cost-based approach, and ended by an application of the simulated-annealing algorithm for systematic computation. Section \ref{section:system} illustrates the modulation strategy of the IPOP-connected DABs. Lastly, the experimental results are laid out in Section \ref{section:experiment}.

\section{Problem Description} 
\label{section:description}

It is common to use multiple modules connected in an IPOP configuration, as illustrated in Fig. \ref{fig:multiple_converter}. In this connection scheme, fully identical modules are typically used to simplify the system's design. Currently, for the optimization of the optimal configuration of identical DC converter sub-modules, two identical IPOP-connected-DAB converters have been proposed \cite{two_DAB}, where equal power sharing results in the highest efficiency.

However, to achieve specific functionalities or operational characteristics, it is also common to use sub-modules with different parameters or topologies \cite{haiyangge2}. This IPOP configuration with non-identical sub-modules brings more flexibility to the system's control strategy, but at the same time, increasing the complexity in system design, such as cost, efficiency, and current stress. This paper primarily focuses on the efficiency optimization problem for such non-identical sub-module IPOP topologies.

\begin{figure}[t]
\centering
\includegraphics[width=0.38\textwidth]{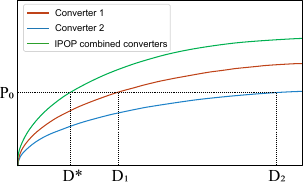}
\caption{\label{fig:two_dab}Power output and duty cycle relationship for non-identical modules.}
\end{figure}

As shown in Fig. \ref{fig:two_dab}, the red and blue lines represent the output power and duty cycle for two different types of non-identical modules, respectively. The green line represents the combined power output when both modules are active. It can be observed that there exists a certain algebraic relationship between power transfer and duty. Under the condition of the same output power, the two different types of modules in the IPOP configuration have different duty cycles. Given a fixed output power ($P_0$), and the IPOP-configured modules operate with a duty cycle of $D^*$, the duty cycle for the red curve (converter 1) is $D_1$, and the duty cycle for the blue curve (converter 2) to achieve the same $P_0$ power output is $D_2$.

\begin{figure}[t]
\centering
\includegraphics[width=0.38\textwidth]{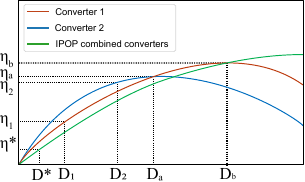}
\caption{\label{fig:duty_eff}Efficiency versus duty cycle for IPOP-connected modules.}
\end{figure}

Fig. \ref{fig:duty_eff} illustrates the relationship between duty cycle and efficiency. It can be observed that under the mentioned duty cycle conditions, the IPOP-connected modules operate at a lower efficiency compared to the cases where the two types of modules are running independently, specifically at the duty cycle $D^*$.

At light load conditions, converter 2 running independently achieves higher efficiency. In the mid-range power output region, converter 1 exhibits higher overall efficiency compared to the other two configurations. However, under high power output conditions, the IPOP configuration with both types of modules achieves the highest overall efficiency, significantly surpassing the other two configurations.

In summary, it can be concluded that the efficiency performance of different module configurations varies in different power ranges. As shown in Fig. \ref{fig:duty_eff}, before the operating point $D_a$, the blue curve (converter 2) exhibits the highest efficiency, while the IPOP configuration with non-identical modules (green curve) shows the lowest. When the duty cycle exceeds $D_a$, the red curve (converter 1) demonstrates significantly higher efficiency compared to the other two modes. When the duty cycle exceeds $D_b$, the efficiency of module 1 and module 2 operating together is higher.

It is possible to optimize the operation of different modules at their respective optimal output power points based on the these characteristics. The IPOP topology with non-identical modules can achieve higher efficiency under high power load conditions. Under light load conditions, each of the two modules can operate on their own efficiency-optimized operating regions, thereby optimizing the combined power converter modules.

Based on the mentioned problem, the objective function for the efficiency optimization problem of the IPOP topology with non-identical modules can be summarized as follows:
\renewcommand{\arraystretch}{1.5} 
\begin{equation*}
\begin{dcases}
max(\eta) = \dfrac{P_{out}}{P_{in}}\\
P_{out} = \sum_{i=1}^{m} P_i(I_i))\\
\end{dcases}\tag{1}
\end{equation*}
The main objective of this study is to maximize the overall efficiency ($\eta$) by optimizing power allocation. The total output power is calculated as the sum of the output powers of each module, which is a function of their respective duty cycles. The proposed solution must satisfy the constraint that each module must operate within its own power range.

This paper introduces a systematic algorithm to calculate the most efficient load distribution scheme, utilizing the given converter's power-current profile. To validate the effectiveness of this algorithm, experiments were conducted using collected data from two different DABs. The Triple Phase Shift (TPS) modulation strategy is employed to minimize current stress under unidirectional power flow, a usual scenario in EV charging applications. Finally, a comprehensive analysis of the system's dynamic characteristics is conducted. The primary objective of this study is to propose an optimized power allocation approach that ensures the highest overall system efficiency while accounting for power range constraints associated with each module. The validity of the proposed algorithm is established through experimentation involving two modules, showcasing a significant enhancement over the conventional approach of equal power distribution among all modules.

\section{Problem Solution}
\label{section:solution}
\subsection{Polynomial fitting of efficiency curves}
Current models on converter losses \cite{transformer_loss, DAB_loss} are able to approximate DC-DC converters with numerous non-idealities, such as copper loss, leakage inductance, eddy current, and hysteresis, etc. The conversion efficiency $\eta$ can be modeled using the quadratic loss model \cite{loss_model}: 
\begin{equation*} \eta=\frac{P_{\text{out}}}{P_{\text{in}}}=\frac{P_{\text{out}}}{P_{\text{out}}+P_{\text{loss}}}=\frac{P_{\text{out}}}{a_{2}\cdot P_{\text{out}}^{2}+(a_{1}+1)\cdot P_{\text{out}}+a_{0}} \tag{2} \end{equation*}
The linear term models the switching loss as a function of $I_{out}$, the quadratic term originates from MOSFETS's resistance loss proportional to $I_{out}^2$, and the constant term models auxiliary circuits. However, non-linear and high-order losses like switching loss are much harder to model. This article, hence, adopts a curve-fitting approach to approximate the input/output power relationship experimentally.

The input and output power can be modeled as $N^{th}$ degree polynomials with respect to current:
\begin{align*} p_{U_1}\left(I \right) =& a_N I^N+a_{N-1}I ^{N-1}+\cdots +a_{2}I + a_1\tag{3}\\ p_{U_2}\left(I \right) =& b_N I ^N+b_{N-1}I ^{N-1}+\cdots +b_{2}I + b_1\tag{4} \end{align*}

Coefficients $\{a_N, a_{N-1}, \cdots, a_1\}$ and $\{b_N, b_{N-1}, \cdots, b_1\}$ can be computed via curve-fitting algorithms. It is worth noted that N is at least 3 given $P_{out} = I^2 R$ to model additional losses, and a higher degree of curve fitting can yield better accuracy in modeling additional high-order non-idealities.

\subsection{Priority assignment to each converter}
The problem of determining the maximum efficiency, which module should operate, and the output power of each designated module belongs to the NP-completeness category: There are no existing polynomial-time algorithms but exhaustive methods. However, given different converters of the same topology but various power ratings, higher-rated power modules are typically less efficient due to their higher operating temperature \cite {thermal} and core loss. Thus, the Priority List (PL) method can be adopted in this case to allocate priority to each converter by arranging each module by their maximum efficiency. Known the expression of $P_{out} = f(P_{in})$, developing an order of the maximum efficiency $\eta$ of each module from greatest to least determines the sequence of which each module should be powered. Each converter will, therefore, join the operation in the order given by the priority queue.

\subsection{Mathematical modeling of N IPOP-connected converters}
With the priority order list computed, the switching point between two operating modes can be determined using equal-incremental cost method. thereby obtaining the optimal combination of modules based on a power output requirement.

\subsubsection{Intersection point solution}
One combination of $m$ converters has the overall efficiency modeled as a function of their individual input and output power relative to their respective current:
\[\eta = \dfrac{P_{out}(I_1) + P_{out}(I_2) + \cdots + P_{out}(I_m)}{P_{in}(I_1) + P_{in}(I_2) + \cdots + P_{in}(I_m)}\tag{5}\]
We can apply Lagrange Multiplier with respect to individual currents.
\[\dfrac{\partial \eta}{\partial I_1} = \dfrac{\dfrac{dP_{out}(I_1)}{dI_1} \sum_{i=1}^{m} P_{in}(I_i) - \dfrac{dP_{in}(I_1)}{dI_1} \sum_{i=1}^{m} P_{out}(I_i)}{(\sum_{i=1}^{m} P_{in}(I_i))^2}\tag{6}\]

\[\dfrac{\partial \eta}{\partial I_2} = \dfrac{\dfrac{dP_{out}(I_2)}{dI_2} \sum_{i=1}^{m} P_{in}(I_i)- \dfrac{dP_{in}(I_2)}{dI_2} \sum_{i=1}^{m} P_{out}(I_i)}{(\sum_{i=1}^{m} P_{in}(I_i))^2}\tag{7}\]

\: \vdots

\[\dfrac{\partial \eta}{\partial I_{m}} = \dfrac{\dfrac{dP_{out}(I_m)}{dI_{m}} \sum_{i=1}^{m} P_{in}(I_{m}) - \dfrac{dP_{in}(I_m)}{dI_{m}} \sum_{i=1}^{m} P_{out}(I_{m})}{(\sum_{i=1}^{m} P_{in}(I_i))^2}\tag{8}\]
At maximum efficiency, all partial derivatives equate to 0. For the $j^{th}$ module:
\[\dfrac{dP_{out}(I_j)}{dI_{j}} \sum_{i=1}^{m} P_{in}(I_{m}) = \dfrac{dP_{in}(I_j)}{dI_{j}} \sum_{i=1}^{m} P_{out}(I_{m})\tag{9}\]

Divide any two equations to cancel common terms:
\[\dfrac{dP_{out}(I_1)}{dP_{in}(I_1)} = \dfrac{dP_{out}(I_2)}{dP_{in}(I_2)} = \cdots = \dfrac{dP_{out}(I_m)}{dP_{in}(I_m)} \tag{10}\]
Hence, as another module joins the operation after switching, equation (9) can be applied to $m+1$ modules.

\[\dfrac{dP_{out}(I'_1)}{dP_{in}(I'_1)} = 
\dfrac{dP_{out}(I'_2)}{dP_{in}(I'_2)} = \cdots = \dfrac{dP_{out}(I'_{m+1})}{dP_{in}(I’_{m+1})} \tag{11}\]

Meanwhile, the total input (12) and output
(13) power are the same for the two combinations at the switching point.

\[\sum_{i=1}^{m} P_{in}(I_i) = \sum_{i=1}^{m+1} P_{in}(I’_i) \tag{12}\]

\[\sum_{i=1}^{m} P_{out}(I_i) = \sum_{i=1}^{m+1} P_{out}(I’_i) \tag{13}\]

Solving equation (10) to (13) yields the input and output power of each module at the intersection point. If a solution is not within the minimum and maximum loads requirement: $P_{out,min}(I_i) \leq P_{out}(I_i) \leq P_{out,max}(I_i)$, according to Karush–Kuhn–Tucker conditions (KKT), the module should be operating at its boundary condition. The load distribution of the remaining modules should be recalculated via equation (10) - (13) by removing the modules operating at the boundary condition from the total output power. Additionally, from equation (10), it can be noted that identical power converter modules always operate at the same output power, which reduces the algorithm complexity.

\subsubsection{Optimal solutions to dynamic power requirements}
Using the Lagrange Multiplier method to solve for $P_{in}(I_i)$ and $P_{out}(I_i)$ for each individual converter requires solving $2m+1$ equations, where there are $m$ different modules available, thus, can be complex for dynamic load adjustment. Additionally, when modules are operating at their boundary conditions, the above-mentioned equations have to be solved repeatedly.
Dynamic optimization algorithms are good candidates for such problems. This article employs the Simulated Annealing Algorithm due to its ability to escape local minimums by generating perturbations on a solution. As the perturbation temperature decreases, the system effectively converges to a globally optimized efficiency. By incorporating the switching point coordinates identified in the preceding section, the optimal solution can be determined by the following algorithm: 
\begin{enumerate}[i)]
\item Define initial temperature $T = T_0$, and the output power of each module is randomly defined based on the given total output power. Obtain the current total efficiency.
\item Calculate the temperature $T = kT$ for the next period. $k \; (0<k<1)$ is the rate of temperature decrease (cooling).
\item Based on the solution obtained, generate a new solution by random disturbance, and the corresponding change in efficiency $\Delta{\eta}$. If $\Delta{\eta}>0$, the new solution is accepted as the optimal solution. Otherwise, Metropolis–Hastings algorithm determines whether to accept the solution by the probability $P = e^{\frac{\Delta{\eta}}{kT}}$.
\item At the current temperature, iterate step 3 by $l$ times.
\item If the temperature has reached the accepted threshold temperature $T_{thres}$, then the current solutions are considered optimal. Otherwise, return to step 2.
\end{enumerate}

\subsubsection{Overall logic diagram}
\begin{figure}
\centering
\includegraphics[width=\linewidth]{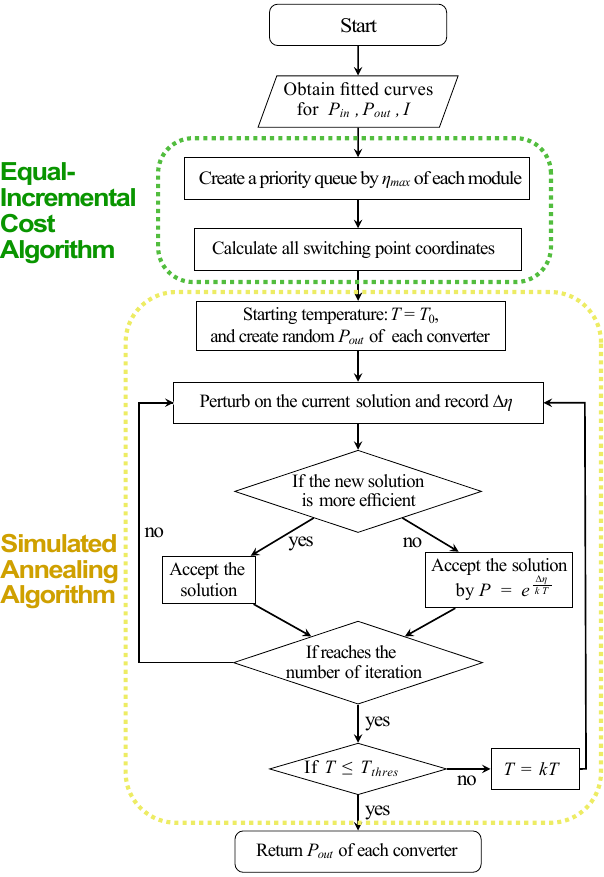}
\caption{\label{fig:flow_chart}Algorithm flow chart diagram, which determines the optimal converter operating combination and power.}
\end{figure}

Therefore, a systematic approach is developed to decide the combination and operating conditions of power converters at maximum overall efficiency. First, a priority queue is computed by sorting the efficiency of each module from greatest to least. Then, the intersection point is computed via the Lagrange Multiplier, which determines the point of switching from one combination to another and the optimal combination between two switching points. Finally, the Simulated Annealing Algorithm decides the optimal solutions to any dynamic output power requirements. 
Therefore, the method proposed by this paper can be summarized by flow chart Fig. \ref{fig:flow_chart}.

\begin{table*}[ht]
  \centering
    \caption{Phase shift strategy for minimized current stress under buck and boost modes \cite{overview}} 
  \begin{tblr}{
	vlines = {},
	hlines = {},
    stretch=1.2,
	cell{2}{1} = {r = 2}{valign = m},
    cell{4}{1} = {r = 2}{valign = m},
    cell{3}{3} = {valign=m},
    cell{2}{3} = {valign=m},
    cell{4}{3} = {valign=m},
    cell{5}{3} = {valign=m},
    colspec={X[-1,c] X[-2,c] X[1,c] Q[0,c] Q[-3,c]},
    rowspec={Q[1, m] Q[5,m] X[5,m] X[5,m] X[5,m]},
}
 Voltage & Range of $p$ & Phase-shift $D_1, D_3$ & Phase-shift $D_2$& Mode\\ 
\vspace{0.2cm} $k>1$\break Boost  & $0\leq p < \dfrac{2(k-1)}{k^2}$ & \vspace{0.2cm}$D_1=1-\sqrt{\dfrac{p}{2(k-1)}}$ $D_3=1-\sqrt{\dfrac{pk^2}{2(k-1)}}$ & $D_2=\dfrac{1}{2}(1+D_1-D_3-\sqrt{-{D_1}^2-{D_3}^2-p+1})$ & 2\\
& $\dfrac{2(k-1)}{k^2}\leq p \leq 1  $ &\vspace{0.2cm} $D_1=(k-1)\sqrt{\dfrac{1-p}{k^2-2k+2}}$ \break$D_3=0$ & $D_2=1-D_3-\sqrt{(1-D_1)(1-D_3)-\dfrac{p}{2}}$& 1\\ \vspace{0.2cm}
$0<k\leq1$\break Buck  & $0\leq p < 2(k-k^2)$ & \vspace{0.2cm}$D_1=1-\sqrt{\dfrac{p}{2k(1-k)}}$ \break $D_3=1-\sqrt{\dfrac{pk}{2(1-k)}}$ &
$D_2=\dfrac{1}{2}(1+D_1-D_3-\sqrt{-{D_1}^2-{D_3}^2-p+1})$ & 2\\
& $2(k-k^2)\leq p \leq 1  $ & \vspace{0.2cm}$D_1=0$ \break $D_3=(1-k)\sqrt{\dfrac{1-p}{2k^2-2k+1}}$ & $D_2=1-D_3-\sqrt{(1-D_1)(1-D_3)-\dfrac{p}{2}}$ & 1\\
\end{tblr}
\label{tab:Table}
\end{table*}

\section{System Modeling}
\label{section:system}
There are generally four efficiency improvement techniques for DABs: Optimization based on power loss model \cite{power_loss_model}, nonactive power optimization \cite{nonactive power}, zero-voltage-switching range optimization \cite{ZVS}, and inductor current optimization \cite{inductor_current}. For a DAB converter, the switching frequency is often restricted due to the physical size and output ripple requirement, therefore cannot be easily changed. However, copper loss, core loss, switching loss, and lumped resistance are functions of the inductor current, making current stress minimization the most effective solution. 

The average output power of a DAB can be modeled as
\[P_{avg}=\dfrac{1}{T}\int_{t_0}^{t_0+T} u_{l}(t)i_l(t)\,dt\tag{14}\]
where $T$ is the switching period, and $u_{l}(t)$ and $i_l(t)$ are the voltage and current on the inductor, which can be controlled by phase shift ratios $D_1$, $D_2$, and $D_3$, where $D_1$ is the primary inner phase shift ratio, $D_2$ is the outer phase shift ratio, and $D_3$ is the secondary inner phase shift ratio. Current phase-shift control method like single-phase-shift (SPS) modulation \cite{SPS}, where $D_1=0$ and $D_2-D_3=0$. However, although SPS control provides certain advantages, it comes with drawbacks such as high circulating power and the absence of Zero Voltage Switching (ZVS) capability. These limitations necessitate the use of higher dimensional modulation techniques \cite{DPS}, such as dual-phase-shift (DPS), extended-phase-shift (EPS), and triple-phase-shift (TPS) that utilizes all three dimensions of control $D_1$, $D_2$, and $D_3$ \cite{overview}, which the experiment will adopt. 

\begin{figure}[h]
\centering
\begin{subfigure}[b]{\linewidth}
\includegraphics[scale=0.148]{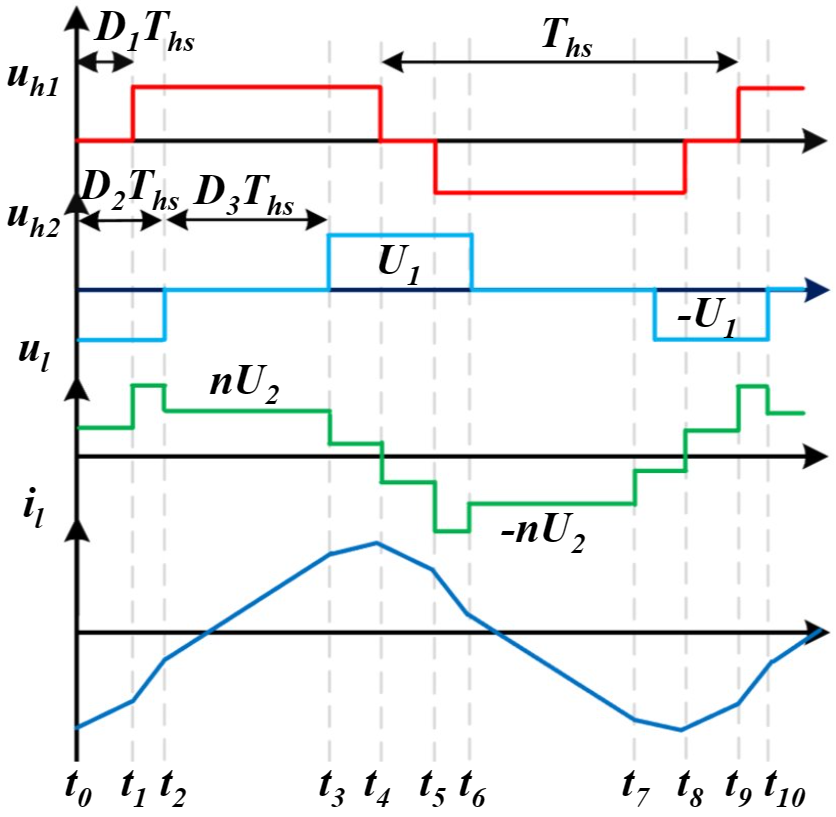}
\includegraphics[scale=0.148]{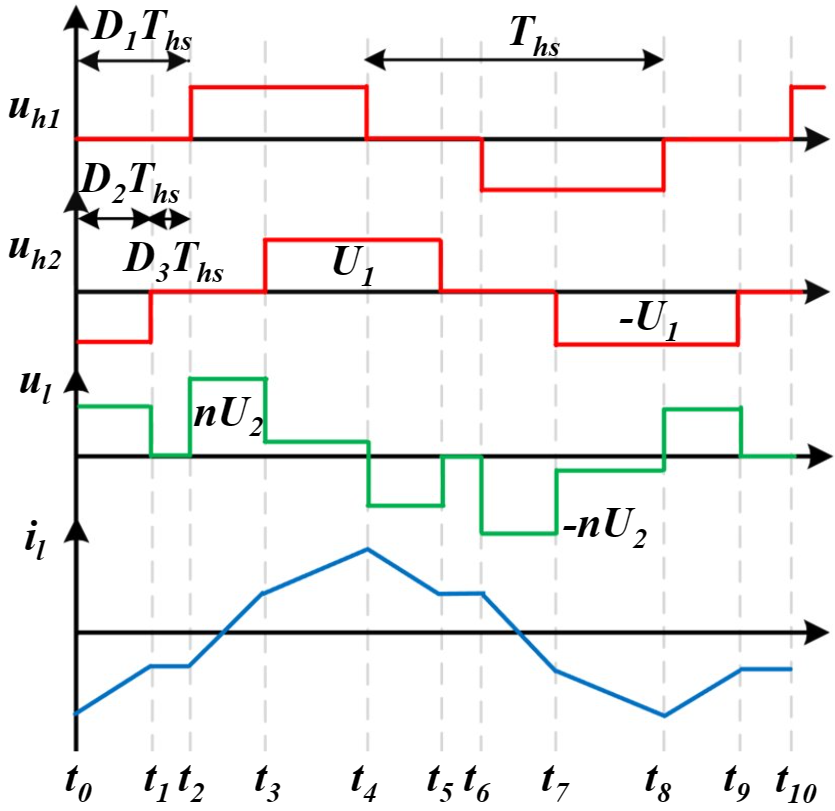}
\end{subfigure}
\caption{I/V characteristics of Mode 1 (left) and 2 (right).}
\label{fig:combined}
\end{figure}

For EV charging applications, maximum forward transmission power is desired. Research \cite{stress_min} lays out 12 modes of TPS control, each corresponding to different forward and reverse charging characteristics. This paper focuses on the two modes as illustrated in table \ref{tab:Table} for optimal forward power transmission with calculated optimal phase-shift values for minimum current stress. Fig. \ref{fig:combined} showcases the inductor currents and voltages corresponding to the two modes. Thus, given the inductor voltage during the whole operating range, the current stress (p. u.) can be illustrated as follows:
\[I_m=\begin{dcases} -kD_1+2D_2+D_3+k-1 & k>1 \\
                     -kD_1+2kD_2-D_3(1-2k)+1-k & 0<k\leq 1
       \end{dcases}\tag{15}\]
\hfill
\[k=\dfrac{n U_{in}}{U_{out}}\tag{16}\]
$n$ is the transformer turns ratio, and $p$ is the transmission power in p.u under the condition $0\leq p \leq 1$.

\section{Experimental Result}
\label{section:experiment}
\begin{figure}
\centering
\includegraphics[width=0.35\textwidth]{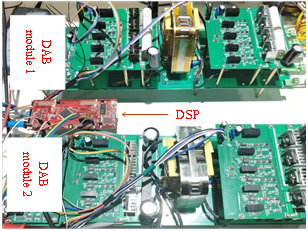}
\caption{\label{fig:exp}Experimental Setup for optimizing DAB Load Distribution.}
\end{figure}

To validate the effectiveness of the proposed method, two experimental prototypes of DAB modules with different parameters were established, as shown in Fig. \ref{fig:exp}. Each DAB module consists of eight Insulated Gate Bipolar Transistors (IGBTs) modal No.IGW60T120. The controller adopted for these modules is TMS320F28335. The total leakage inductance of the high-frequency isolation transformer and the sum of auxiliary inductances are $100uH$ and $150uH$ respectively. Detailed experimental parameters for the DAB modules are provided in table \ref{tab:experiment_setup} in the appendix.

\begin{figure}
         \centering
      \setkeys{Gin}{width=\linewidth}
     \begin{subfigure}[t]{0.347\textwidth}
              \centering
         \includegraphics{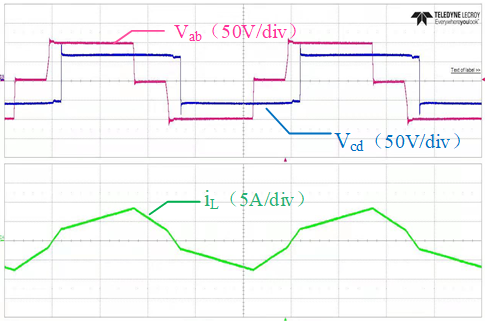}
         \caption{100uH}
     \end{subfigure}
     \hfill
     \begin{subfigure}[t]{0.347\textwidth}
         \centering
         \includegraphics{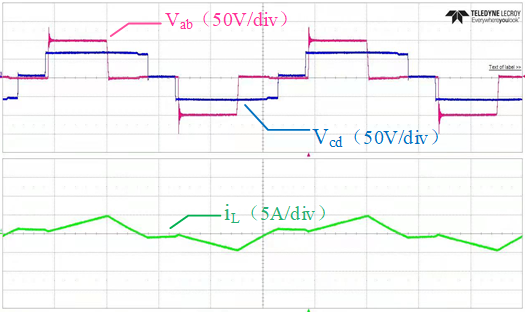}
         \caption{150uH}
     \end{subfigure}
        \caption{Voltage and current waveform for optimal current stress for two DABs}
        \label{fig:combined2}
    \end{figure}

All DAB modules adopt TPS modulation strategy \cite{overview} to reduce switching losses and optimize the current stress in each module. As illustrated in the experimental waveform Fig. \ref{fig:combined2}, under the buck operation with a voltage range of 100-80V and a transmission power of 600W, the modulation strategy minimizes the current stress. For Module 1 ($100uH$) and Module 2 ($150uH$), the waveforms of the high-frequency transformer's voltage and inductor current are shown. This modulation strategy ensures that the inductor current stress is minimized for this specific working condition. By employing the TPS modulation strategy, the DAB module achieves efficient and optimized performance.

\begin{figure}
\centering
\includegraphics[width=0.43\textwidth]{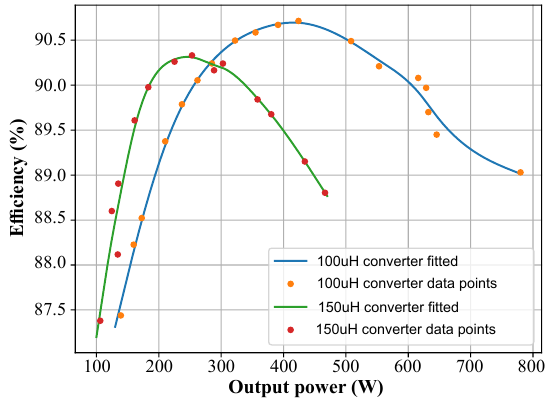}
\caption{\label{fig:exp_fit}Experimental data and fitted-fitting result for the two DAB modules}
\end{figure}

To validate the algorithm proposed in this research, power and efficiency data points were collected from two DAB modules. A curve-fitting approach is applied to model the system behavior in all load conditions. As shown in Fig. \ref{fig:exp_fit}, the module with 150uH leakage inductance exhibits higher efficiency under light-load conditions, whereas the module with 100uH leakage inductance performs better at output powers exceeding 290W. 

\begin{figure}[H]
\centering
\includegraphics[width=0.43\textwidth]{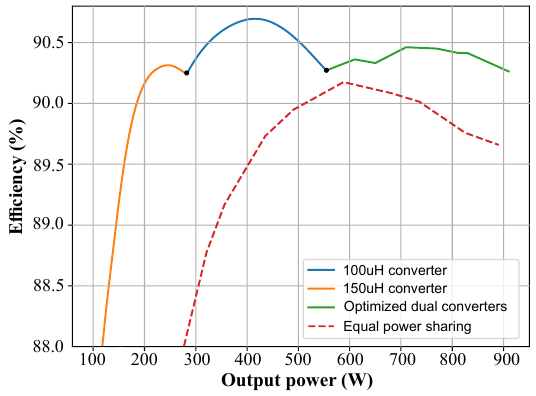}
\caption{\label{fig:3_exp}Overall operating scheme with experiment data and compare group}
\end{figure}

The algorithm developed in this research generates the optimal load distribution scheme, as shown in Fig. \ref{fig:3_exp}. The plot illustrates the efficiency of the two DAB modules, and the experimental results closely align with the theoretical expectations. The switching points are identified at $p=282.75W$ and $p=549.80W$ according to Equal-Incremental Cost algorithm. In addition, the experiment includes a control group where two modules operate at the same power level for comparison purposes. The experiment's results showcase a substantial increase in efficiency, with improvements of up to 0.74\% compared to the even load distribution scheme. This significant enhancement further emphasizes the algorithm's superiority and enables a significantly wider output power range, outperforming individual modules and expanding its applicability across various converter topologies.
\section{Conclusion}
IPOP-connected converters are cost-effective solutions to wide power requirement for EV charging applications. This paper demonstrates that smaller converters can achieve a significant boost in efficiency under the proposed algorithm.  

This article proposed a systematic way to compute the operating condition of each module using collected data points. A curve-fitting approach is utilized to define the efficiency curves experimentally. To locate the switching point between two modes in multi-IPOP-connected systems, this study adopts the Equal Incremental Rate method. The system calculations are performed using the Simulated Annealing algorithm to obtain the optimal output distribution scheme for each module under dynamic power requirements. Finally, this study is validated using the two prototype DABs. Experimental results demonstrate that under the operating condition $p=282.75W$ and $p=549.80W$, the system should switch from 150uH to 100uH, and 100uH to two modules, when both modules are assigned load distribution according to the Equal Marginal Rate method, a significant efficiency improvement is achieved. By comparison with the two modules operating at the same output power, the proposed algorithm can achieve an increase in efficiency by 0.74\%.



\appendix
\begin{appendices}

See Table \ref{tab:experiment_setup}.

\begin{table}[H]
\caption{\label{tab:experiment_setup} Experiment and simulation parameters}
\begin{center}
\begin{tabular}{ |c|c|c|c| } 
 \hline
 Parameter & Symbol & Value & Unit \\ 
  \hline
 Input voltage & $U_1$ & 100 & $V$\\ 
  \hline
 Output voltage & $U_2$ & 80 & $V$\\ 
  \hline
 Switching frequency & $fs$ & 10 & $kHz$\\
  \hline
 Inductance of module 1 & $Lr_1$ & 100 & $uH$\\
  \hline
 Inductance of module 2 & $Lr_2$ & 150 & $uH$\\
  \hline
 Input capacitance & $C_1$ & 320 & $uF$\\
  \hline
 Output capacitance & $C_2$ & 320 & $uF$\\
 \hline
 MOSFET's gate resistance & $R_{gint}$ & 4 & $\Omega$\\
  \hline
 Transformer turns ratio & n & 1:1 & -\\
 \hline
\end{tabular}
\end{center}
\end{table}
\end{appendices}




\ifCLASSOPTIONcaptionsoff
  \newpage
\fi



%
\bibliographystyle{unsrt}

%








\end{document}